\documentclass[12pt, referee]{iopart}
\usepackage{iopams}  
\usepackage{graphicx}

\begin{document}


\title[
Calculation of electron-impact excitation parameters
]
{
Methods, algorithms and computer codes
for calculation of electron-impact excitation parameters
}

\author{P. Bogdanovich, R. Kisielius and D. Stonys}

\address{Institute of Theoretical Physics and Astronomy, Vilnius University, 
A. Go\v stauto 12,  LT-01108, Vilnius, Lithuania}


\begin{abstract}

We describe the computer codes, developed at Vilnius University, for the 
calculation of electron-impact excitation cross sections, collision strengths, 
and excitation rates in the plane-wave Born approximation. These codes 
utilize the multireference atomic wavefunctions which are also adopted to 
calculate radiative transition parameters of complex many-electron ions. 
This leads to consistent data sets suitable in plasma modelling codes. 
Two versions of electron scattering codes are considered in the present work, 
both of them employing configuration interaction method for inclusion of 
correlation effects and Breit-Pauli approximation to account for relativistic 
effects. These versions differ only by one-electron radial orbitals, where the 
first one employs the non-relativistic numerical radial orbitals, while another 
version uses the quasirelativistic radial orbitals. The accuracy of produced 
results is assessed by comparing radiative transition and electron-impact 
excitation data for neutral hydrogen, helium and lithium atoms as well as
highly-charged tungsten ions with theoretical and experimental data available 
from other sources. \\



\end{abstract}

\maketitle

\section{Introduction}
\label{intro}

In modelling both high temperature plasma (stellar atmosphere, nuclear fusion) 
and low temperature plasma, such as planetary nebulae, working material of 
spectroscopic and medical devices, one needs data on free-electron interaction 
with atoms and ions. For the consistency of plasma models, it is highly 
desirable that such data are calculated within the same approximation, using 
the identical multireference atomic wavefunctions, applying the same methods to 
include relativistic and correlation corrections as it has been done in
production of spectroscopic data, such as energy levels, oscillator strengths
and radiative transition probabilities. 

Over many years, original methods and computer codes designated to calculate
various spectral parameters of atoms and ions have been developed in the 
Department of Atomic Theory, Institute of Theoretical Physics and Astronomy, 
Vilnius University \cite{1, 2}. Currently these codes were supplemented with 
our new codes computing electron-impact excitation parameters for ions in the 
plane-wave Born approximation. The main purpose of this development is to 
establish a consistent and complete set of data, necessary for plasma spectra 
modelling. Such a set will be suitable for our newly-developed database ADAMANT 
(Applicable DAta of Many-electron Atom eNergies and Transitions), where the 
main requirement is to produce data sets within the same atomic wavefunctions 
base, hence, simplifying an application of these data in modelling codes. 
In Section~\ref{method} of the present work, for the first time we describe 
the implemented calculation methods. Further, in Section~\ref{algorithm},
we describe the algorithms implemented in our computer codes.

Developed methods and computer codes are equiped to calculate many-electron
atoms and ions with open s-, p-, d- and f- shells when consistent inclusion of 
the correlation effects is necessary. It is important to perform calculation of
the electron-impact excitation parameters using an extensive configuration 
interaction basis in order to match them to other spectroscopic parameters,
such as energy levels, transition probabilities, oscillator strengths) 
determined in the same approximation and similar accuracy. These codes will be 
employed in cases when the adaption of other more accurate theoretical mathods, 
such as R-matrix approximation (RM) or converged close-couling approximation 
(CCC)), is very  difficult or even impossible due to complex atomic structure
(e.g. for heavy multicharged ions). The plane-wave Born approximation to 
calculate the electron-impact excitation parameters was chosen as a suitable one
be the developers of Atomic Data and Analysis System (ADAS), see \cite{32}. ADAS
needs the data for complex tungsten ions which can not be determined using other
approximations.

In the current work, in order to benchmark the adopted approximations and 
developed codes, we present the investigation of the electron-impact excitation 
cross sections for light atoms, namely H, He, and Li. There is substantial 
amount of atomic structure and electron-atom interaction data for these atoms, 
both theoretical and experimental ones. Most of them can be found in the NIST 
Atomic Spectra Database \cite{10}, the NIST Electron-Impact Excitation Cross 
Sections Database \cite{4}, and the CCC database \cite{11}. We choose H, He, and
Li atoms for comparison since these are the only atoms having electron-impact
excitation data in the NIST database \cite{4} which usually contains only very
reliable parameters. We have performed our calculation by employing both the 
non-relativistic radial orbitals and the quasirelativistic ones. Determined 
results are compared in Section~\ref{result}. Conclusions are 
presented in the final Section.

\section{Description of adopted method}
\label{method}

The total cross section $\sigma$, describing the interaction of an incident 
electron having the energy $\varepsilon$ with an atom, for the excitation from 
the energy level $K_0\lambda_0J_0$ to the level $K_1\lambda_1J_1$ is expressed 
as a sum of the excitation cross sections with different ranks $\kappa$:

\begin{equation}
\fl \sigma(K_0\lambda_0J_0,K_1\lambda_1J_1,\varepsilon) = \sum_{\kappa}
\sigma_{\kappa}(K_0\lambda_0J_0,K_1\lambda_1J_1,\varepsilon)\,.
\label{sigma}
\end{equation}
Here $K$ denotes electronic configuration, $\lambda$ denotes level number, 
$J$ stands for the total angular momentum and $\varepsilon$ is the energy of 
incident electron. Contrary to the case of radiative transitions, 
$\sigma_{\kappa}$ does not contain fine-structure parameter $\alpha$, and the 
summation in \eref{sigma} must be performed over all possible ranks $\kappa$. 
This summation must be performed over even or odd  $\kappa$ values, 
depending on the parity of the initial and final levels, including $\kappa = 0$,
if the excitation process does not change their parity. The rank $\kappa$ must 
satisfy the triangular condition with the even perimeter $(J_0,\kappa,J_1)$. 

The excitation cross section $\sigma_{\kappa}$ of any rank  $\kappa$ is 
expressed using the matrix element $\mathcal{S}_{\kappa}$ of electron-impact 
excitation operator:

\begin{equation}
\fl \sigma_{\kappa}(K_0\lambda_0J_0,K_1\lambda_1J_1,\varepsilon) =
\frac{4\pi a_{0}^2}{\varepsilon}(2J_1 + 1)
\int\limits_{k_0-k}^{k_0+k}\left| \mathcal{S}_{\kappa}
(K_0\lambda_0J_0,K_1\lambda_1J_1,q) \right|^2 \frac{dq}{q^3}\,.
\label{sigma_k}
\end{equation}
Here $a_0 = 0.5291772 \times 10^{-10}$\,m is atomic length unit. We must 
underline that all our calculations are performed in the system of atomic units.
Integration is performed over the difference of the electron momenta; $k_0$ is 
the momentum of an incident electron, and $k$ is the momentum of an outgoing 
electron.

Our developed methods and computer codes determine cross sections for 
transitions between different energy levels by employing multiconfigurational 
multi-term wavefunctions in $LSJ$-coupling:

\begin{equation}
\fl \Psi(K_0\lambda_0J_0|x) = \sum_{KTLS}
a(K_0\lambda_0J_0,KTLSJ_0) \Psi(KTLSJ_0|x)\,.
\label{ksi}
\end{equation}
Here $TLS$ denote the intermediate and final orbital and spin-angular momenta 
for the pure $LSJ$-coupling wavefunction, and $x$ denotes coordinates of all 
electrons of the wavefunction $\Psi$. In this case, the matrix element of
the electron-impact excitation operator is expressed by relation

\begin{eqnarray}
\fl \mathcal{S}_{\kappa}(K_0\lambda_0J_0,K_1\lambda_1J_1,q) & = 
\sum_{KTLS,K^{\prime}T^{\prime}L^{\prime}}
a(K_0\lambda_0J_0,KTLSJ_0) \nonumber \\
& \times \mathcal{S}_{\kappa}(KTLSJ_0,K^{\prime}T^{\prime}L^{\prime}SJ_1,q) 
a(K^{\prime}T^{\prime}L^{\prime}SJ_1,K_1\lambda_1J_1) \,.
\label{sk}
\end{eqnarray}

In this approximation, the excitation process does not change the term 
multiplicity $S$. The matrix elements for the excitation process in pure 
$LSJ$-coupling can be described by a product

\begin{equation}
\fl \mathcal{S}_{\kappa}(KTLSJ_0,K^{\prime}T^{\prime}L^{\prime}SJ_1,q) =
Q_{\kappa}(KTLS,K^{\prime}T^{\prime}L^{\prime}S) W_{\kappa}(J_0,J_1)
R_{\kappa}(n_{\mathrm i}\ell_{\mathrm i},n_{\mathrm f}\ell_{\mathrm f},q)\,.
\label{sk1}
\end{equation}

Here $Q_{\kappa}(KTLS,K^{\prime}T^{\prime}L^{\prime}S)$ is an angular integral 
for the matrix element of the electric multipole transition operator of the rank
$\kappa$, which can be determined using code \cite{5} developed for the 
calculation of the radiative transition matrix elements. Therefore, the rank 
$\kappa$ must satisfy the triangular condition with even perimeter 
$(L_0,\kappa,L_1)$. The expression for $Q_{\kappa}$ also includes sub-matrix 
elements of the spherical function, consequently, an additional triangular 
condition of with the even parameter $(\ell_{\mathrm i},\kappa,\ell_{\mathrm f})$
arises, where $\ell_{\mathrm i}$ and $\ell_{\mathrm f}$ are the orbital momenta 
of excited electron before and after collision. The coefficient describing the 
bounding of the total orbital momentum $L$ and the total spin $S$ into the total 
angular momentum $J$ is the same one as in calculation of the radiative 
transition matrix element:

\begin{equation}
\fl W_{\kappa}(J_0,J_1) = (-1)^{L^{\prime}+\kappa-J-S}
\left\{ \begin{array}{c c c}
L         & L^{\prime}& \kappa \\
J^{\prime}& J         & S      \\
\end{array}\right\}\,,
\label{wk}
\end{equation}
where the element in curly brackets denotes the $6j$-coefficient. 
$R_{\kappa}(n_{\mathrm i}\ell_{\mathrm i},n_{\mathrm f}\ell_{\mathrm f},q)$ is 
a radial integral:

\begin{equation}
\fl
R_{\kappa}(n_{\mathrm i}\ell_{\mathrm i},n_{\mathrm f}\ell_{\mathrm f},q)
= \int\limits_0^{\infty}P(n_{\mathrm i}\ell_{\mathrm i}|r) 
P(n_{\mathrm f}\ell_{\mathrm f}|r)
[j_{\kappa}(qr) - \delta_{\kappa0}]dr\, ,
\label{rk}
\end{equation}
where $P(n\ell|r)$ denotes one-electron radial orbital of the initial and the
final state of an electron, $\delta_ {\kappa0}$ is a Kronecker delta function, 
and $j_{\kappa}(qr)$ is a spherical Bessel function \cite{6}:

\begin{equation}
\fl
j_{\kappa}(qr) = \left(\frac{\pi}{2qr} \right)^{1/2} \,
J_{\kappa+1/2}(qr)\,.
\label{jk}
\end{equation}

Finally, we can define an expression for the total electron-impact excitation 
cross section: 

\begin{eqnarray}
\fl 
\sigma(K_0\lambda_0J_0,K_1\lambda_1J_1,\varepsilon)& = \nonumber \\
&\frac{4\pi a_0^2}{\varepsilon}\,(2J_1 + 1) \sum_{\kappa}\int\limits_{k_0-k}^{k_0+k} 
\sum_{KTLS,K^{\prime}T^{\prime}L^{\prime}}\bigg [ 
\,a(K_0\lambda_0J_0,KTLSJ_0) \nonumber \\ 
& \times Q_{\kappa}(KTLS,K^{\prime}T^{\prime}L^{\prime}S) 
W_{\kappa}(J_0,J_1) \nonumber \\
& \times \int\limits_0^{\infty}P(n_{\mathrm i}\ell_{\mathrm i}|r)
P(n_{\mathrm f}\ell_{\mathrm f}|r) [j_{\kappa}(qr) - 
\delta_{\kappa0}]\, dr \nonumber \\
& \times a(K^{\prime}T^{\prime}L^{\prime}SJ_1,K_1\lambda_1LS,) \bigg ]^2 
\frac{dq}{q^3}\, .
\label{sigma2}
\end{eqnarray}

From \eref{sigma2} one can see that the expression for electron-impact 
excitation cross section has a similar structure and analogous angular 
integrals compared to calculation of the radiative transition probabilities.
Nevertheless, the radial part in the electron-impact excitation cross section 
expression is much more complex. There is a double integral and sum over all 
possible ranks $\kappa$ instead of one single integral. Moreover, calculations 
must be performed for various energies of the incident electron $\varepsilon$. 
These energies are usually expressed using the excitation threshold energies, 
which are determined as a difference between the energies of the final and 
initial levels. Consequently, there is a different set of free-electron 
energies for each excitation channel. Therefore, a special attention should be 
paid in order to simplify and speed-up calculation of the radial integrals for 
the excitation cross sections.

\section{Calculation algorithm}
\label{algorithm}

At first, we determine all necessary angular integrals 
$Q_{\kappa}(KTLS,K^{\prime}T^{\prime}L^{\prime}S)$ for the matrix element
of transition operator. In order to avoid repetitive calculations of the 
radial integrals with close values of the parameter $q$, we exploit 
interpolation of radial integrals. For this purpose, we determine the minimum 
and maximum values of $q$ from the initial information about the investigated 
excitation. Within these limits, a grid of 1001 $q$-values is generated. 
Since the functions in calculated integral are more sensitive at the lower 
parameter values, this grid of $q$-values is formed using a logarithmic step.
Next, a two-dimensional grid of $q$ and the radial variable $r$ products is 
generated. A grid for the Bessel function \eref{jk} is determined according to 
the grid of the parameter $q \cdot r$ for all necessary rank $\kappa$ values, 
starting with $\kappa =0$. After the Bessel function grid is generated, all the 
possible integrals \eref{rk} are determined for given $q$ values. These 
computations do not require too much time even if the quantities of the 
parameter $q$ and radial variable $r$ values are large. 

After the $R_{\kappa}$ grid is determined, electron-impact excitation cross 
sections can be produced. In most cases, the incident electron energies 
$\varepsilon$ are introduced in the excitation threshold units. Nevertheless, 
there is a possibility in our code to define a different energy grid.  A grid of
the parameter $q$ is determined for each incident electron energy in order to 
perform integration in \eref{sigma_k}. As in previous case, this grid has a 
logarithmic step. We have performed a computational experiment and have 
determined, that a desirable accuracy can be achieved if the Simpson's rule 
\cite{7} is applied for the integral when a number of points is as low as 
25. The integrals 
$R_{\kappa}(n_{\mathrm i}\ell_{\mathrm i},n_{\mathrm f}\ell_{\mathrm f},q)$, 
which are necessary for 
calculation of the electron-impact excitation matrix elements in \eref{sk1},
are determined by interpolating their values according to the previously 
generated values. Only the nearest four points are required for this 
interpolation. Integration over the parameter $q$ is performed for the 
determined matrix element. Further, the calculated 
$\sigma_{\kappa}(K_0\lambda_0J_0,K_1\lambda_1J_1,\varepsilon)$ values are 
summed over all possible ranks $\kappa$ as in \eref{sigma}, and the total 
electron-impact excitation cross section is determined. We are performing 
parallel computing for the cross sections of different ranks $\kappa$ in order 
to accelerate production of results.

The above described calculation method is realized in two computer codes. 
One of these codes exploits the non-relativistic Hartree-Fock (HF) radial 
orbitals. Another one employs the numerical quasirelativistic (QR) radial 
orbitals. The performance of our new codes was tested on complex heavy ions, 
such as Hf -- Hf$^{3+}$, Ta$^{+}$ -- Ta$^{4+}$, W$^{2+}$ -- W$^{5+}$, and 
Re$^{3+}$ -- Re$^{6+}$. Such complex systems require to deal correctly with both 
relativistic corrections and correlation effects. Unfortunately, there are no
reliable experimental or theoretical data for such complex systems published so 
far. Furthermore, in order to benchmark these newly developed codes, we have 
performed calculation of the electron-impact excitation cross sections for 
three light neutral atoms, namely, hydrogen, helium and lithium. Some data for
the W$^{45+}$ ion were also calculated and compared with calculations performed
using more sophisticated R-matrix method in the next section. Finally, we
demonstrate abilities of our computer codes by presenting the electron-impact
excitation cross sections for several low-ionization stages of tungsten ions.

\section{Results and discussions} 
\label{result}

The NIST database \cite{4} contains the total electron-impact cross sections for 
the transitions from the ground configuration with one level to an excited 
configuration rather than to separate levels. Our computer codes are designed 
to determine cross sections for the transitions between the individual levels 
of many-electron atoms and ions \eref{sigma2}. Therefore the results determined
in the present work were summed over all levels of the final configuration. 
In general case, a cross section averaging is performed by applying 
expression:	

\begin{equation}
\fl \sigma(K_0,K_1,\varepsilon) = 
\frac{\sum_{\lambda_{0i}J_{0i},\lambda_{1j}J_{1j}}(2J_{0i} + 1)
\sigma(K_0\lambda_{0i}J_{0i},K_1\lambda_{1j}J_{1j},\varepsilon)}
{\sum_{\lambda_{0i}J_{0i}}(2J_{0i} + 1)} \, .
\label{skk1}
\end{equation}

The most of the data for comparison with our results are taken from \cite{8}. 
The electron-impact excitation cross sections in that work are scaled 
($BE$-scaling) according to a method developed for neutral atoms in \cite{9}:

\begin{equation}
\fl
\sigma_{BE}(K_0,K_1,\varepsilon) = \frac{\varepsilon}{\varepsilon +B +E}
\, \sigma(K_0,K_1,\varepsilon) \, .
\label{sigmaBE}
\end{equation}

Here $B$ is the ionization energy of the initial level, $E$ is the transition 
energy. Such a scaling significantly alters cross section values, particularly 
those at the energies close to the excitation threshold. This leads to a 
substantially better agreement between theoretical results and experimental 
data. Since the scaled data for cross sections are presented in \cite{8}, 
we utilize the $BE$-scaling for all our results in the current work. Likewise 
in \cite{8}, we utilize the $B$ values from \cite{4}, and these values agree 
favorably with our theoretical results. For the transition energy $E$, we have 
utilized our {\it ab initio} energy values. The electron-impact excitation 
cross sections in \cite{8} were computed using the one-configuration Dirac-Fock
approximation. Therefore, in order to improve the accuracy of calculated data, 
authors introduced an additional $f$-scaling:

\begin{equation}
\fl
\sigma_{BE,f}(K_0,K_1,\varepsilon) =
\frac{f_{\mathrm{ac}}}{f_{\mathrm{sc}}} \sigma_{BE}(K_0,K_1,\varepsilon)\, ,
\label{sigmaBEF}
\end{equation}
where the ratio of oscillator strengths was applied. Here $f_{\mathrm{sc}}$ is 
a theoretical (one-configuration approximation) oscillator strength, and
$f_{\mathrm{ac}}$ is a high-accuracy oscillator strength from \cite{10}. 
Since the multiconfiguration approximation is employed in the present work, 
the $f$-scaling is not performed for production of our results.

For comparison with the results from \cite{8}, we introduce a mean-square 
deviation $MSD$ defined as:

\begin{equation}
\fl
MSD = \sqrt{\frac{1}{N}\sum_{n=1}^N
\left[ \frac{\sigma_{BE,f}(K_0,K_1,\varepsilon_n)-\sigma_{BE}(K_0,K_1,\varepsilon_n)}
{\sigma_{BE,f}(K_0,K_1,\varepsilon_n)} 
\right]^2} \times 100\% \, ,
\label{msd}
\end{equation}
where $\sigma_{BE,f}(K_0,K_1,\varepsilon_n)$ stands for the data from \cite{8}, 
$\sigma_{BE}(K_0,K_1,\varepsilon_n)$  is our calculated cross section results. 
Summations were performed for all electron energies $\varepsilon_n$ presented 
in the NIST database.

\subsection{Excitation of hydrogen atom}
\label{4.1}

\noindent

\Table{\label{t1}
The electron-impact excitation cross sections $\sigma$ (in 10$^{-17}$ cm$^2$) 
for H atom at various incident electron energies $\varepsilon$. }
\br
\multicolumn{1}{l}{$\varepsilon$}&
\multicolumn{1}{c}{HF}&
\multicolumn{1}{c}{${BE,f}$}&
\multicolumn{1}{c}{CCC}&
\multicolumn{1}{c}{Exp}\\
\multicolumn{1}{l}{(eV)}&
\multicolumn{1}{c}{}&
\multicolumn{1}{c}{\cite{8}}&
\multicolumn{1}{c}{\cite{11}}&
\multicolumn{1}{c}{\cite{12}}\\
\mr
\multicolumn{5}{c}{1s--2p}\\
14  & 3.506& 3.512& 3.849&     \\
15  & 3.920& 3.926& 4.021& 4.8 \\
20  & 5.272& 5.278& 5.163& 5.7 \\
30  & 6.312& 6.319& 6.187& 6.4 \\
40  & 6.524& 6.532&      & 6.7 \\
45  & 6.506& 6.513& 6.499&     \\
70  & 6.025& 6.032& 6.142&     \\
100 & 5.330& 5.337& 5.485&     \\
150 & 4.417& 4.423& 4.582&     \\
200 & 3.772& 3.777& 3.912&     \\
500 & 2.076& 2.081& 2.126&     \\
1000& 1.242& 1.246& 1.261&     \\
\multicolumn{5}{c}{1s--3p}\\
14  & 0.437& 0.438& 0.564&      \\
15  & 0.536& 0.537& 0.587&      \\
20  & 0.827& 0.828& 0.752&      \\
30  & 1.039& 1.040& 0.960&      \\
45  & 1.086& 1.087& 1.024&      \\
70  & 1.010& 1.012& 1.014&      \\
100 & 0.895& 0.896& 0.899&      \\
150 & 0.741& 0.742& 0.761&      \\
200 & 0.632& 0.633& 0.652&      \\
500 & 0.346& 0.347& 0.355&      \\
\multicolumn{5}{c}{1s--4p}\\
14  & 0.127& 0.127& 0.203&      \\
15  & 0.169& 0.170& 0.209&      \\
20  & 0.284& 0.284& 0.242&      \\
30  & 0.364& 0.364& 0.325&      \\
45  & 0.382& 0.383& 0.352&      \\
70  & 0.357& 0.357& 0.355&      \\
100 & 0.316& 0.316& 0.315&      \\
150 & 0.262& 0.262& 0.268&      \\
200 & 0.223& 0.223& 0.230&      \\
500 & 0.122& 0.122& 0.125&      \\
\br
\endTable

We have investigated the electron-impact excitation of the 1s electron to the 
2p, 3p, and 4p states. As one can expect, an agreement with data from \cite{8} 
is excellent for all three transitions within the complete energy range. Both 
the HF and the QR results display $MSD=0.22\%$ for these three transitions, and
our results are slightly lower than the data from \cite{8}. The deviations for 
the 1s--2p excitation are larger at the beginning of the investigated 
energy range, where the deviations are approximately $0.43\%$ at 
$\varepsilon = 11$\,eV, and at the end of the presented energy range, where the 
deviations are roughly $0.82\%$ at $\varepsilon = 3000$\,eV. In the middle of 
the energy range, the deviations do not exceed $0.12\%$. A completely similar 
situation is for the excitation to the 3p state. For excitation to the 4p state,
the deviations at the low and at the high energy end increase, but this increase
does not change $MSD$ noticeably.

In \tref{t1} we present the electron-impact excitation cross sections of 
hydrogen atom for the cases, where it is possible to make comparison with the
data produced by several authors. For our data, we present only the HF results 
because, as one can expect, the difference between our HF and QR results 
appears only in the fifth or sixth significant digit. We present the theoretical
results, determined using the convergent close-coupling method (CCC) \cite{11} 
and the experimental data from \cite{12}. It is not unusual that our data agree
better with the data from \cite{8} rather than with those from \cite{11}. Their
agreement with the experimental data from \cite{12} is within error limits.

\subsection{Excitation of helium atom}
\label{4.2}

We determined the cross sections of electron-impact excitation from the 1s shell
to the 2p and 3p shells. Our HF results were obtained in a following way. 
First of all, the Hartree-Fock equations were solved for the 1s2s configuration
of helium atom. At next step, the equations for the 2p, 3s, and 3p electrons 
were solved in a potential of the frozen 1s electron. The determined radial 
orbital basis was complemented with the transformed radial orbitals (TRO) 
\cite{2,13}, which described virtual electron excitations and had the principal
quantum number values $4 \leq n \leq 9$ and all possible values of the orbital 
momentum $\ell$. Consistent with our other calculations, we employed the same 
radial orbital basis both for the even-parity and odd-parity configurations. 
Therefore, we can avoid problems related to the non-orthogonality of radial 
orbitals. 

The correlation effects are included within the configuration interaction (CI) 
approach. We selected admixed configurations which had averaged weights, 
calculated in the second order of perturbation theory \cite{12}, larger than 
$10^{-10}$ in the wavefunction expansions of the adjusted configurations 1s2s, 
1s2p and 1s3p by applying the method described in \cite{14,15}. We can adopt 
such a minute selection criteria, because our investigated configurations have 
only two electrons. Therefore, the constructed Hamiltonian matrices are 
relatively small. We apply Breit-Pauli approximation to include relativistic 
effects. Furthermore, the parameters of radiative transitions and 
electron-impact excitations are calculated after the eigenvalues and the 
eigenfunctions are determined. Our QR+CI approximation results are obtained in 
very similar way, but this time the quasirelativistic equations \cite{16,17,18}
are solved instead of Hartree-Fock equations in order to determine the 
one-electron radial orbitals.

\Table{\label{t2}
Energy levels $E$ and radiative transition probabilities $A$ in He.}
\br
\multicolumn{1}{c}{Level}&
\multicolumn{3}{c}{$E$ (cm$^{-1}$)}&
\multicolumn{3}{c}{$A$ (s$^{-1}$)}\\
\multicolumn{1}{c}{}&
\crule{3}&
\crule{3}\\
&
\multicolumn{1}{c}{\cite{19}}&
\multicolumn{1}{c}{HF+CI}&
\multicolumn{1}{c}{QR+CI}&
\multicolumn{1}{c}{\cite{20}}&
\multicolumn{1}{c}{HF+CI}&
\multicolumn{1}{c}{QR+CI}\\
\hline
1s2p $^3P_1$& 169087& 169031& 169106& 1.764E+2& 1.586E+2& 1.714E+2\\
1s2p $^1P_1$& 171135& 171078& 171140& 1.799E+9& 1.794E+9& 1.826E+9\\
1s3p $^3P_1$& 185565& 185513& 185573&         & 3.093E+1& 3.312E+1\\
1s3p $^1P_1$& 186209& 186160& 186255& 5.663E+8& 5.604E+8& 5.637E+8\\
\br
\endTable

The investigated energy levels and their radiative dipole emission transition 
probabilities to the ground level are given in \tref{t2}. We present only those
levels which can be excited by electron impact from the ground level via the 
dipole transitions. For the He atom, our calculated results are compared 
with those from the NIST database. Those data were determined using 
high-accuracy calculations for the helium isotopes \cite{19}. It is evident 
from \tref{t2} that our theoretical level energies agree very well with NIST 
data. The deviations reach only few one-tenths of cm$^{-1}$. Although the 
relativistic effects for the helium atom are small, our QR+CI calculations 
give more accurate energy level values. Similar high accuracy is evident for 
the emission transition probabilities, when they are compared with the data 
from compilation \cite{20}. The accuracy of the QR+CI results is fine even for
the intercombination transition 1s2p $^3$P$_1 - $ 1s$^2\,\,^1$S$_0$.

Our cross sections agree well with the data from \cite{8} for the 1s--2p 
excitation. Some larger deviations, reaching few percent, are noticeable only 
for the low incident electron energies $\varepsilon$.  Otherwise, these 
deviations are significantly lower than $1\%$. Consequently, the mean-square 
deviations from all data contained in the NIST database \cite{8} are $1.2\%$ 
for the HF+CI approximation, and $MSD = 2.6\%$ for the QR+CI approximation. 
The larger $MSD$ value for the quasirelativistic case is caused by more than 
two-times larger  deviations, compared to the HF+CI approximation, at low 
electron energies ($\varepsilon < 50$\,eV). The deviations from the compiled 
data \cite{8} both for the QR+CI and for the HF+CI results are of similar size 
at higher electron energies.


\Table{\label{t3}
The 1s--2p electron-impact excitation cross sections $\sigma$ (in 10$^{-18}$ 
cm$^2$) for He atom at various incident electron energies $\varepsilon$.}
\br
\multicolumn{1}{l}{$\varepsilon$}&
\multicolumn{1}{c}{HF+CI}&
\multicolumn{1}{c}{QR+CI}&
\multicolumn{1}{c}{$BE,f$}&
\multicolumn{2}{c}{CCC}&
\multicolumn{3}{c}{Exp}&
\multicolumn{1}{c}{RV}\\
&  &&&
\crule{2}&
\crule{3}&
& \\
\multicolumn{1}{l}{(eV)}&
\multicolumn{1}{c}{}&
\multicolumn{1}{c}{}&
\multicolumn{1}{c}{\cite{8}}&
\multicolumn{1}{c}{\cite{11}}&
\multicolumn{1}{c}{\cite{21}}&
\multicolumn{1}{c}{\cite{22}}&
\multicolumn{1}{c}{\cite{23}}&
\multicolumn{1}{c}{\cite{21}}&
\multicolumn{1}{c}{\cite{25}}\\
\mr

25   & 2.851 & 2.733 &  2.933 &  2.23 &     &       &  1.165&     &   1.2\\ 
30   & 4.566 & 4.440 &  4.669 &  3.89 &     &   3.75&  3.408&     &   3.7\\ 
35   & 5.770 & 5.664 &  5.875 &       &     &   5.43&       &     &      \\ 
50   & 7.806 & 7.775 &  7.903 &  8.19 &     &   8.20&  8.370&     &   8.5\\ 
60   & 8.452 & 8.455 &  8.542 &       &     &   9.52&  9.240&     &   9.5\\ 
70   & 8.799 & 8.827 &  8.883 &       &     &   9.77&  9.660&     &  10.0\\ 
80   & 8.965 & 9.009 &  9.043 & 10.60 &     &  10.15&  9.810&     &  10.2\\ 
90   & 9.014 & 9.071 &  9.088 &       &     &  10.15&  9.820&     &  10.2\\ 
100  & 8.989 & 9.055 &  9.060 & 10.87 &     &  10.10&  9.740&     &  10.1\\ 
120  & 8.812 & 8.889 &  8.876 &       &     &   9.63&       &     &      \\ 
150  & 8.406 & 8.490 &  8.462 &       &     &   9.18&  8.780&     &   9.2\\ 
180  & 7.961 & 8.047 &  8.012 &       &     &   8.81&       &     &      \\ 
200  & 7.670 & 7.756 &  7.718 &  9.05 & 8.97&   8.30&  7.747&  9.2&   8.1\\ 
250  & 7.004 & 7.087 &  7.046 &       &     &   7.63&       &     &      \\ 
300  & 6.433 & 6.512 &  6.471 &       &     &   6.95&       &     &      \\ 
350  & 5.947 & 6.022 &  5.982 &       &     &   6.41&       &     &      \\ 
400  & 5.532 & 5.603 &  5.564 &       &     &   5.94&       &     &      \\ 
500  & 4.862 & 4.926 &  4.890 &  5.54 & 5.54&   5.07&  4.579&  5.2&      \\ 
900  & 3.330 & 3.376 &  3.350 &  3.70 &     &       &       &     &      \\ 
1000 & 3.096 & 3.139 &  3.116 &       &     &   3.14&  2.903&     &      \\ 
1500 & 2.316 & 2.349 &  2.333 &       &     &   2.34&       &     &      \\ 
2000 & 1.869 & 1.896 &  1.885 &       &     &   1.85&  1.747&     &      \\ 
\br
\endTable

We compare our results with the data from other authors in \tref{t3}. 
As in \tref{t1}, we display only those electron energies, for which there
are data obtained using different methods. Such are the theoretical results 
from \cite{11,21}, the experimental data from \cite{21,22,23}, and the 
recommended values (RV) from \cite{25}. One can see from \tref{t3} that our 
results are closer to those from \cite{8} than to other theoretical data. 
This can be explained by the fact that we adopt the same scaling procedure 
\eref{sigmaBE} as in \cite{8}. Unfortunately, we can not achieve close agreement
to the recommended values (RV) \cite{25}, especially at low electron energies.


\Table{\label{t4}
The 1s--3p electron-impact excitation cross sections (in 10$^{-18}$ cm$^2$) for 
He atom at various incident electron energies $\varepsilon$.}
\br
\multicolumn{1}{l}{$\varepsilon$}&
\multicolumn{1}{c}{HF+CI}&
\multicolumn{1}{c}{QR+CI}&
\multicolumn{1}{c}{$BE,f$}&
\multicolumn{1}{c}{CCC}&
\multicolumn{2}{c}{Exp}&
\multicolumn{1}{c}{RV}\\
&&&&&
\crule{2}& \\
\multicolumn{1}{l}{(eV)}&
\multicolumn{1}{c}{}&
\multicolumn{1}{c}{}&
\multicolumn{1}{c}{\cite{8}}&
\multicolumn{1}{c}{\cite{11}}&
\multicolumn{1}{c}{\cite{22}}&
\multicolumn{1}{c}{\cite{23}}&
\multicolumn{1}{c}{\cite{26}}\\
\mr
25  & 0.509& 0.494& 0.498& 0.40&     &      &      \\
30  & 1.028& 0.992& 1.002& 0.75& 0.74& 0.582&      \\
35  & 1.366& 1.313& 1.329&     & 1.11&      &      \\
40  & 1.611& 1.548& 1.565& 1.40& 1.38& 1.400&      \\
45  & 1.793& 1.726& 1.740&     & 1.66&      &      \\
50  & 1.930& 1.863& 1.871& 1.88& 1.84& 2.166&      \\
60  & 2.112& 2.052& 2.043&     & 2.16& 2.098&      \\
70  & 2.214& 2.165& 2.138&     & 2.34& 2.216&      \\
80  & 2.266& 2.228& 2.185& 2.46& 2.43& 2.265&      \\
90  & 2.287& 2.259& 2.202&     & 2.45& 2.274&      \\
100 & 2.288& 2.267& 2.199& 2.59& 2.44& 2.258&      \\
120 & 2.252& 2.245& 2.161&     & 2.41&      &      \\
150 & 2.156& 2.162& 2.066&     & 2.26& 2.041&      \\
180 & 2.046& 2.060& 1.959&     & 2.20&      &      \\
200 & 1.973& 1.990& 1.888& 2.23& 2.08& 1.805& 2.08 \\
250 & 1.804& 1.825& 1.726&     & 1.88&      &      \\
300 & 1.658& 1.680& 1.587&     & 1.77&      &      \\
350 & 1.532& 1.554& 1.468&     & 1.63&      &      \\
400 & 1.425& 1.447& 1.365&     & 1.51&      &      \\
500 & 1.251& 1.272& 1.201& 1.37& 1.29& 1.086& 1.29 \\
900 & 0.854& 0.868& 0.823& 0.92&     &      &      \\
1000& 0.793& 0.807& 0.766&     & 0.79& 0.691&      \\
1500& 0.592& 0.602& 0.573&     & 0.59&      &      \\
2000& 0.476& 0.484& 0.463&     & 0.47& 0.417&      \\
\br
\endTable

For the 1s--3p excitation, our results agree with those from \cite{8} slightly
worse compared to the 1s--2p excitation results. Here $MSD = 3.7\%$ in the 
HF approach, and $MSD = 4.2\%$ in the quasirelativistic approach. It is 
interesting that agreement for this transition is better at lower energies, 
but it exceeds $4\%$ and $5\%$, correspondingly, at high electron energies. 
Our HF+CI cross section values are slightly larger than those from \cite{8},
whereas the QR+CI data are slightly smaller at low energies, but they become 
larger at higher electron energies. The electron-impact excitation cross
sections are tabulated in \tref{t4}, where we present the theoretical data 
(CCC) from \cite{11}, the experimental data from \cite{22,23}, and the 
recommended values from \cite{26}.  The fact that our calculated values are 
larger than those from \cite{8} brings them closer to the recommended data 
from \cite{26}.

\subsection{Excitation of lithium atom}
\label{4.3}

\Table{\label{t5}
Energy levels $E$ and radiative transition probabilities $A$ of lithium atom.}
\br
\multicolumn{1}{c}{}&
\multicolumn{3}{c}{$E$ (cm$^{-1}$)}&
\multicolumn{3}{c}{$A$ (s$^{-1}$)}\\
&
\crule{3}&
\crule{3}\\
\multicolumn{1}{c}{Level}&
\multicolumn{1}{c}{NIST\,\cite{10}}&
\multicolumn{1}{c}{HF+CI}&
\multicolumn{1}{c}{QR+CI}&
\multicolumn{1}{c}{NIST\,\cite{20}}&
\multicolumn{1}{c}{HF+CI}&
\multicolumn{1}{c}{QR+CI}\\
\mr
2p $^2P_{1/2}$& 14904& 14923& 14900& 3.689E+7& 3.719E+7& 3.694E+7\\
2p $^2P_{3/2}$& 14904& 14923& 14900& 3.689E+7& 3.719E+7& 3.694E+7\\
3p $^2P_{1/2}$& 30925& 30924& 30915& 1.002E+6& 0.943E+6& 0.945E+6\\
3p $^2P_{3/2}$& 30925& 30924& 30915& 1.002E+6& 0.943E+6& 0.945E+6\\
\br
\endTable

For the lithium atom, we investigated the excitation of the 2s electron to the
2p and 3p shells. Calculations were performed in the following way. 
First of all, the Hartree-Fock equations were solved for the 1s$^2$2s 
configuration. Afterward, the equations for the 2p, 3s, 3p, and 3d electrons 
were solved in a frozen-core potential. This basis of radial orbitals was 
complemented with the TRO describing virtual electron excitations for the 
principal quantum number $4 \leq n \leq 11$ and for all possible values of the 
orbital quantum number $\ell$.

We selected all admixed configurations which have their averaged weights 
\cite{12} larger than $10^{-15}$ in the adjusted configurations 1s$^2$2s, 
1s$^2$2p, and 1s$^2$3p. Calculations in the quasirelativistic approximation 
were performed in a completely analogous way. Such an extension of the TRO basis
and the reduction of the configuration selection parameter down to $10^{-15}$, 
compared to the calculation of He atoms described in Sect.\ref{4.2}, was 
necessary in order to ensure the convergence of the 3p--2s transition 
probability. The determined energy levels and the transition probabilities $A$ 
are presented in \tref{t5}. One can see that agreement of the energy level 
values with data from the NIST database is rather good. Agreement of the 2p--2s
transition probability values with the compilation data from \cite{20} is fine, 
too. For the 3p--2s transition, we cannot achieve such an agreement, and the 
discrepancy of our data is approximately $6\%$.


\Table{\label{t6}
The 2s--2p electron-impact excitation cross sections $\sigma$ (in 10$^{-16}$ 
cm$^2$) for Li atom at various incident electron energies $\varepsilon$.}
\br
\multicolumn{1}{l}{$\varepsilon$}&
\multicolumn{1}{c}{HF+CI}&
\multicolumn{1}{c}{QR+CI}&
\multicolumn{1}{c}{$BE,f$}&
\multicolumn{3}{c}{Exp}&
\multicolumn{1}{c}{CCC}\\
&&&&
\crule{3}&
\\
\multicolumn{1}{l}{(eV)}&
\multicolumn{1}{c}{}&
\multicolumn{1}{c}{}&
\multicolumn{1}{c}{\cite{8}}&
\multicolumn{1}{c}{\cite{27}}&
\multicolumn{1}{c}{\cite{28}}&
\multicolumn{1}{c}{\cite{29}}&
\multicolumn{1}{c}{\cite{30}}\\
\mr
 2.1  & 18.459& 18.618& 18.374&      & 13.020&    &      \\
 2.3  & 23.744& 23.884& 23.591&      & 17.595&    &      \\
 2.7  & 30.236& 30.360& 30.017&      & 26.920&    &      \\
 3.5  & 36.959& 37.065& 36.691&      & 35.717&    &      \\
 4.0  & 39.258& 39.356& 38.978&      & 38.796&    & 36.08\\
 5.0  & 41.762& 41.844& 41.474&      & 41.084&    &      \\
 5.4  & 42.284& 42.360& 41.995&      &       &  49&      \\
 6.0  & 42.738& 42.807& 42.451&      &       &    & 39.70\\
 6.6  & 42.908& 42.971& 42.623&      & 41.172&    &      \\
 8.0  & 42.611& 42.662& 42.333&      &       &    & 39.71\\
 10.0 & 41.335& 41.373& 41.072& 38.00&       &  44& 38.33\\
 10.81& 40.689& 40.724& 40.433&      & 38.972&    &      \\
 15.0 & 37.106& 37.126& 36.877&      &       &    & 34.47\\
 15.64& 36.570& 36.588& 36.346&      & 35.365&    &      \\
 20.0 & 33.183& 33.194& 32.984& 33.10&       &  36& 31.36\\
 23.78& 30.655& 30.662& 30.473&      & 30.351&    &      \\
 25.0 & 29.915& 29.921& 29.738&      &       &    & 28.87\\
 30.0 & 27.221& 27.222& 27.061&      &       &    & 26.57\\
 38.6 & 23.592& 23.591& 23.457&      & 23.691&    &      \\
 40.0 & 23.096& 23.094& 22.963&      &       &    & 22.78\\
 50.0 & 20.105& 20.101& 19.991&      &       &    & 20.09\\
 60.0 & 17.839& 17.835& 17.739& 17.50&       &  28&      \\
 63.56& 17.160& 17.154& 17.064&      & 17.410&    &      \\
 70.0 & 16.062& 16.057& 15.973&      &       &    & 16.35\\
 99.15& 12.532& 12.527& 12.465&      & 12.765&    &      \\
 100.0& 12.454& 12.449& 12.387& 12.40&       &    & 12.90\\
 149.4&  9.207&  9.202&  9.160&      &  9.387&    &      \\
 150.0&  9.179&  9.174&  9.131&  9.63&       &    &  9.60\\
 200.0&  7.333&  7.328&  7.296&  7.56&       &    &  7.66\\
 249.9&  6.139&  6.135&  6.110&      &  6.236&    &      \\
 300.0&  5.296&  5.292&  5.271&      &       &    &  5.56\\
 400.0&  4.183&  4.179&  4.164&      &       &    &  4.34\\
 400.5&  4.178&  4.175&  4.160&      &  4.230&    &      \\
 600.0&  2.982&  2.979&  2.970&      &       &    &  3.08\\
 601.4&  2.976&  2.973&  2.965&      &  3.010&    &      \\
 800.0&  2.337&  2.335&  2.330&      &       &    &  2.43\\
 802.3&  2.332&  2.330&  2.324&      &  2.348&    &      \\
1000.0&  1.932&  1.930&  1.927&      &       &    &  2.01\\
1404.2&  1.442&  1.441&  1.440&      &  1.447&    &      \\
2000.0&  1.061&  1.060&  1.061&      &       &    &  1.08\\
\br
\endTable
      
The cross sections for the electron-impact excitation of the 2s--2p transition 
are presented in \tref{t6}. One can see that our results completely agree 
with the data from \cite{8}. For the most of calculated energies, the deviations 
from the NIST database values are less than $0.5\%$. For the HF+CI calculations,
$MSD = 0.55\%$, and for the QR+CI calculations, $MSD = 0.66\%$. Their agreement 
with the experimental data from \cite{27,28,29} is fine, too. Some larger 
deviations appear at the electron energies lower than 3 eV. Deviations from the
theoretical data \cite{30} are more noticeable. The reason for this discrepancy
is that the $BE,f$-reduction procedure have not been applied in \cite{30}.


\Table{\label{t7}
The 2s--3p electron-impact excitation cross sections $\sigma$ 
(in 10$^{-17}$ cm$^2$) for Li atom at various incident electron energies 
$\varepsilon$.}
\br
\multicolumn{1}{l}{$\varepsilon$}&
\multicolumn{1}{l}{HF+CI}&
\multicolumn{1}{l}{QR+CI}&
\multicolumn{1}{l}{$BE,f$}\\
\multicolumn{1}{l}{(eV)}&
\multicolumn{1}{l}{}&
\multicolumn{1}{l}{}&
\multicolumn{1}{l}{\cite{8}}\\
\mr
4    & 3.470& 3.486&  5.329\\
5    & 6.313& 6.317&  9.753\\
6    & 6.481& 6.481& 10.059\\
7    & 6.256& 6.254&  9.740\\
8    & 5.952& 5.950&  9.289\\
9    & 5.647& 5.645&  8.826\\
10   & 5.361& 5.359&  8.387\\
15   & 4.258& 4.256&  6.667\\
20   & 3.537& 3.535&  5.530\\
30   & 2.656& 2.655&  4.137\\
40   & 2.136& 2.135&  3.314\\
60   & 1.545& 1.544&  2.382\\
80   & 1.216& 1.215&  1.866\\
100  & 1.005& 1.004&  1.536\\
120  & 0.858& 0.858&  1.308\\
140  & 0.750& 0.749&  1.139\\
160  & 0.666& 0.666&  1.010\\
180  & 0.600& 0.600&  0.908\\
200  & 0.546& 0.546&  0.825\\
250  & 0.446& 0.446&  0.672\\
300  & 0.378& 0.378&  0.568\\
400  & 0.291& 0.291&  0.434\\
600  & 0.200& 0.200&  0.297\\
800  & 0.153& 0.153&  0.227\\
1000 & 0.125& 0.124&  0.184\\
1500 & 0.085& 0.085&  0.125\\
2000 & 0.065& 0.065&  0.095\\
3000 & 0.044& 0.044&  0.065\\
\br
\endTable

A completely different situation is observed for the 2s--3p excitation by 
electron impact. The cross sections of this process are presented in \tref{t7}. 
Our HF+CI results agree with the  QR+CI results very well within all 
investigated electron energy range. Meanwhile, they are approximately by $34\%$
smaller when compared to the data from \cite{8}. It is important to notice that
the 3p--2s radiative transition probability $A$ values can change significantly, 
depending on the composition of the radial orbital basis and the amount of 
admixed configurations. Our calculated values converge to the data 
presented in \tref{t5} only after the CI expansion basis is made very large. 
At the same time, the electron-impact excitation cross section values have been
rather stable and practically have not changed following the increase of 
the wavefunction basis, after all the most important admixed configurations 
have been included. 

We do not apply the $f$-reduction \eref{sigmaBEF} for our data because the 
difference between our radiative transition probabilities and the data from 
\cite{8} is only $6\%$, and the transition energies agree really well. 
Therefore, that reduction can not alter significantly our calculated 
electron-impact excitation cross section values. The overestimated cross 
section $\sigma_{BE,f}$ values in \cite{8} can be a consequence of the 
$f$-scaling procedure applied in that work. This is caused by the situation, 
where the E1 transition probability is proportional to the matrix element of 
the $<r^1>$ operator which is more important at longer distances. Meanwhile, 
when calculating the electron-impact excitation cross sections, one needs to 
apply the $<f_1(qr)>$ operator which has a maximum located closer to a nucleus.

Unfortunately, we could not find any additional references on the experimental
data or the calculation results of the 2s--3p excitation of Li atoms.

\subsection{Excitation of tungsten ions}
\label{4.4}

Previous comparisons for H, He, and Li demonstrate that the applied method is 
able to produce good-quality results for light neutral atoms. Described 
computer codes are mainly targeted for the production of electron scattering 
parameters for substantially more complex many-electron ions where it is very 
difficult to apply more accurate approximations and calculation methods 
developed for the electron-ion collision processes. In such a case, it is not 
a simple task even to determine correct and inclusive-enough CI wavefunction 
expansion for the considered target system. The problems increase even more 
when one has to consider target+electron system containing significantly more 
configurations. This is the main reason why there are not so many data 
determined in R-matrix (RM) or convergent close-coupling (CCC) approach for 
heavy ions. Therefore application of rather simple approximation, such as 
plane-wave Born for the scattering process description becomes the most optimal,
if not the only possible way so far to determine necessary data.

\begin{figure}
\includegraphics[scale=0.6]{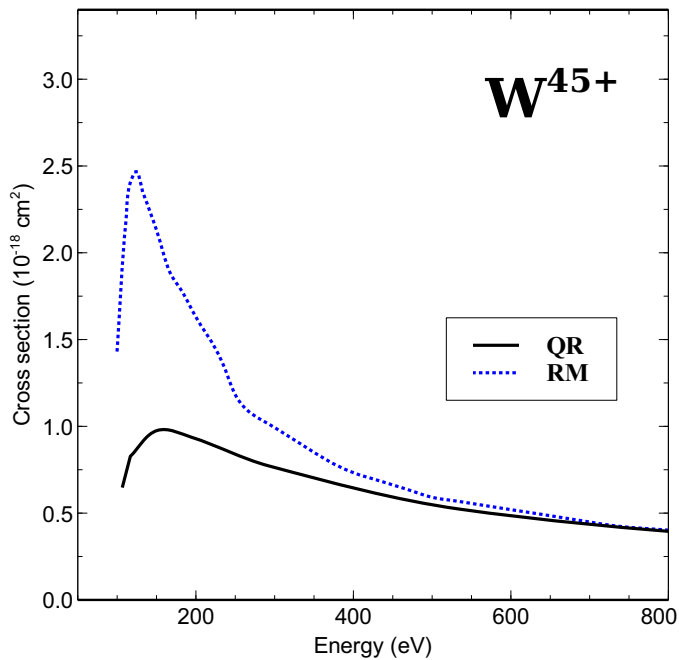}
\caption{\label{w45} Comparison of the $W^{45+}$ electron-impact excitation 
cross sections from the ground state to the 4p $^2$P$_{1/2}$ level.
The dotted curve is the R-matrix (RM) damped cross section convoluted with a 
30~eV Gaussian from \cite{31}, the solid curve represent our 
quasirelativistic (QR) calculation.
}
\end{figure}

There are just few data sources of collision data for complex many-electron 
tungsten ions. In \cite{31} the results of fully relativistic, radiatively 
damped R-matrix calculations for W$^{44+}$ and W$^{45+}$ are presented. Authors 
in \cite{31} were able to perform such kind of scattering calculations for 
these ions because of relatively small number of levels in the target expansion.
We compare data from \cite{31} to our calculation for the W$^{45+}$ in 
Fig.~\ref{w45}. The threshold energy for this transition $\Delta E = 97.275$ eV 
agrees very well with data from \cite{31} where $\Delta E = 97.233$ eV. 
Similarly, the radiative transition probability in \cite{31} 
$A = 5.07 \times 10^{10}$ s$^{-1}$ is very close to our determined value of 
$A = 5.04 \times 10^{10}$ s$^{-1}$ (the deviation is just $0.6\%$). 

One can clearly see two different trends in that comparison. At the 
higher-energy end ($E > 300$ eV), the deviations are reasonably small, and they 
differ just by a few percent when the incident electron energy is above 800 eV. 
At the low incoming electron energies ($E < 300$ eV), the difference between our
data and R-matrix values is significant. The R-matrix calculation data are some 
2.5 times larger than our cross section values at $E=125$ eV, where cross 
sections reach maximum values. But that difference falls sharply when the 
incident electron energy increases. There are two main reasons for such a 
behaviour. The R-matrix electron-impact excitation cross-section contains
abundance of resonances which reflect excitation through autoionizing levels of
the W$^{44+}$ ion. These resonances are located near excitation threshold, 
therefore they can make up $20\% - 25\%$ of the convoluted cross section value,
as one can see from figure 4 in \cite{31}. Another reason is that there is a
strong interaction between incident free electron and the bound electrons of the 
W$^{45+}$ ion. This interaction cannot be included in the plane-wave 
approximation adopted in the current work. Such an omission leads to smaller 
cross section values in our calculation. It is necessary to underline, that 
quite a large electron energy range is required when collision rates are 
determined. That leads to a rather small difference (especially at higher 
temperatures) in determined rate values applied in high-temperature plasma 
modelling.

Even more complicated calculations are required when low ionization stages of
tungsten ions are investigated. We have performed electron-impact excitation
cross section calculations for several isoelectronic sequences such as
Tm-, Yb-, Lu-, Hf- like ions including W$^{2+}$ -- W$^{5+}$ ions. Data for these
ions are important in fusion plasma spectra modelling. We have determined both
the spectroscopic parameters (energy levels, transition wavelengths, radiative 
transition probabilities) and the electron-impact excitation parameters 
(collision strengths, cross sections, collision rates) for several ions of these
sequnces. Our data are incorporated in Atomic Data and Analysis System (ADAS)
\cite{32} as the basic parameters and are utilized to determine the derived 
parameters for various plasma conditions.

\begin{figure}
\includegraphics[scale=0.5]{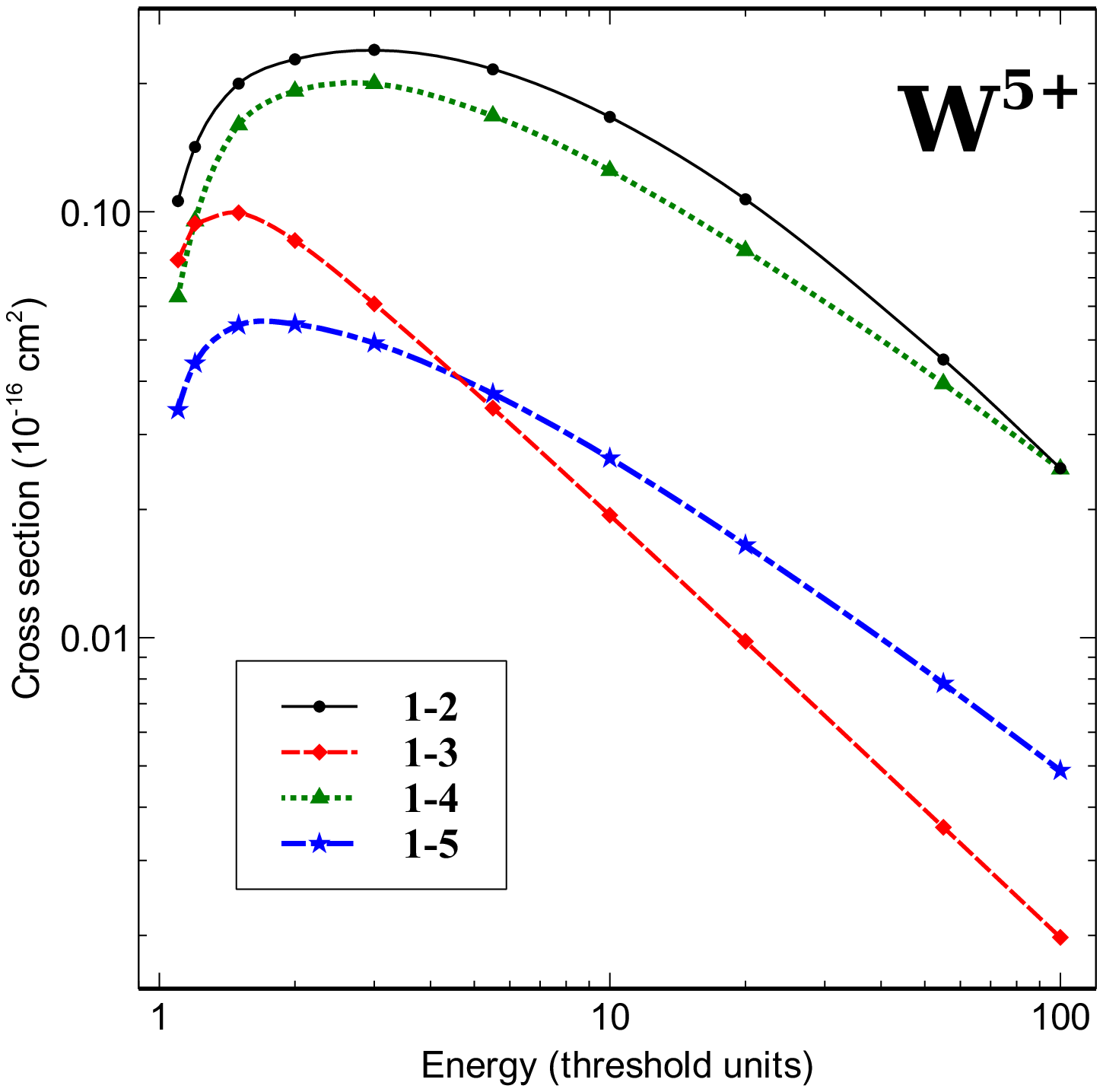}
\caption{\label{w5_1} 
Electron-impact excitation cross sections (in $10^{-16}$ cm$^2$) 
from the ground state 
5p$^6$5d\,\,$^2$D$_{3/2}$ 
of the W$^{5+}$ ion to the excited states
5p$^6$5d\,\,$^2$D$_{5/2}$ (solid curve), 
5p$^6$6s\,\,$^2$S$_{1/2}$ (dashed curve),
5p$^6$6p\,\,$^2$P$_{1/2}$ (dotted curve) and 
5p$^6$6p\,\,$^2$P$_{3/2}$ (dash-dotted curve). 
Incident electron energies are given in transition threshold units. 
$X=E/\Delta E$.
}
\end{figure}

\begin{figure}
\includegraphics[scale=0.5]{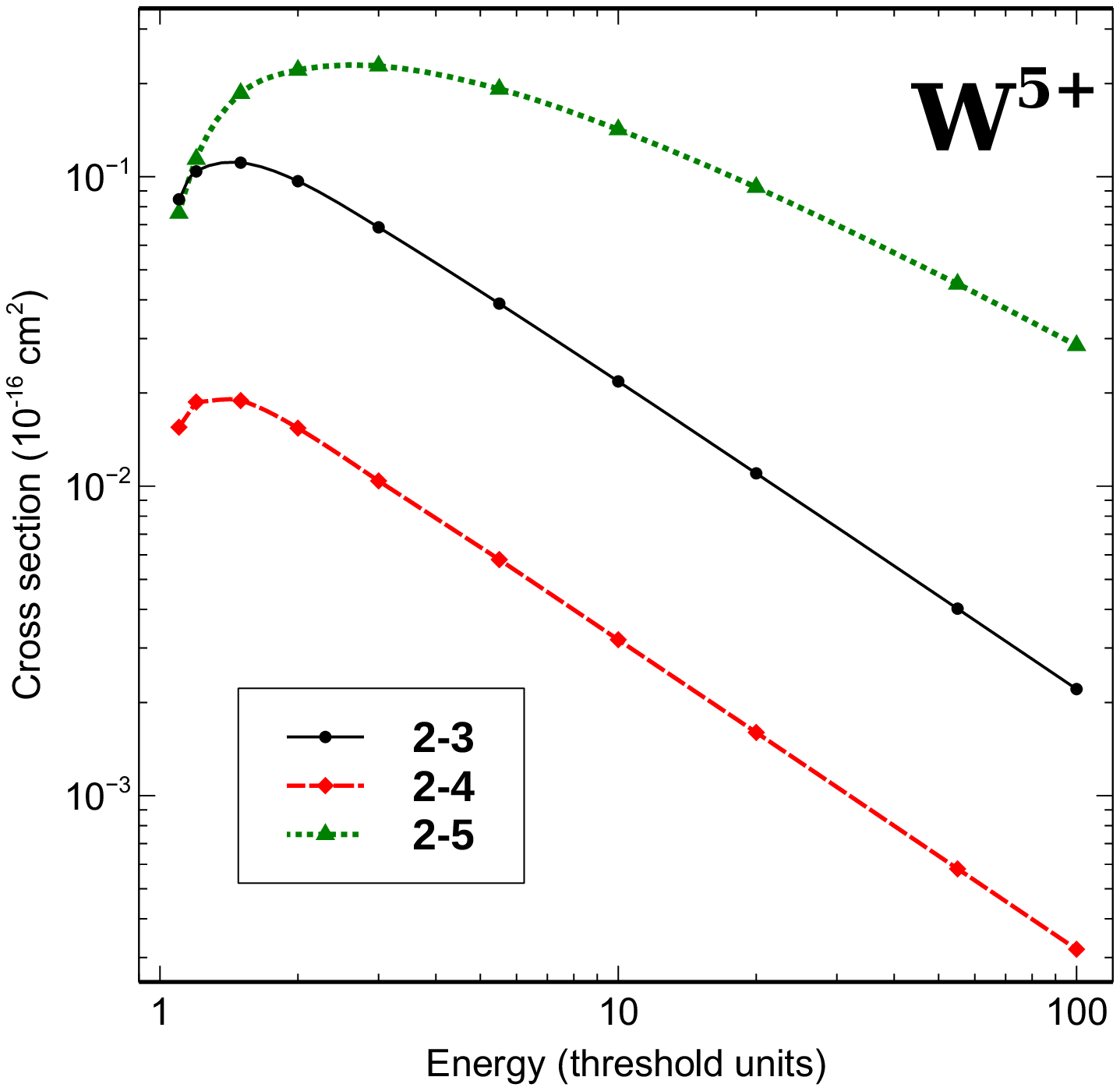}
\caption{\label{w5_2} 
Electron-impact excitation cross sections (in $10^{-16}$ cm$^2$) 
from the first excited state 
5p$^6$5d\,\,$^2$D$_{5/2}$ 
of the W$^{5+}$ ion to the excited states
5p$^6$6s\,\,$^2$S$_{1/2}$ (solid curve),
5p$^6$6p\,\,$^2$P$_{1/2}$ (dashed curve) and 
5p$^6$6p\,\,$^2$P$_{3/2}$ (dotted curve).
Incident electron energies are given in transition threshold units. 
}
\end{figure}

\begin{figure}
\includegraphics[scale=0.5]{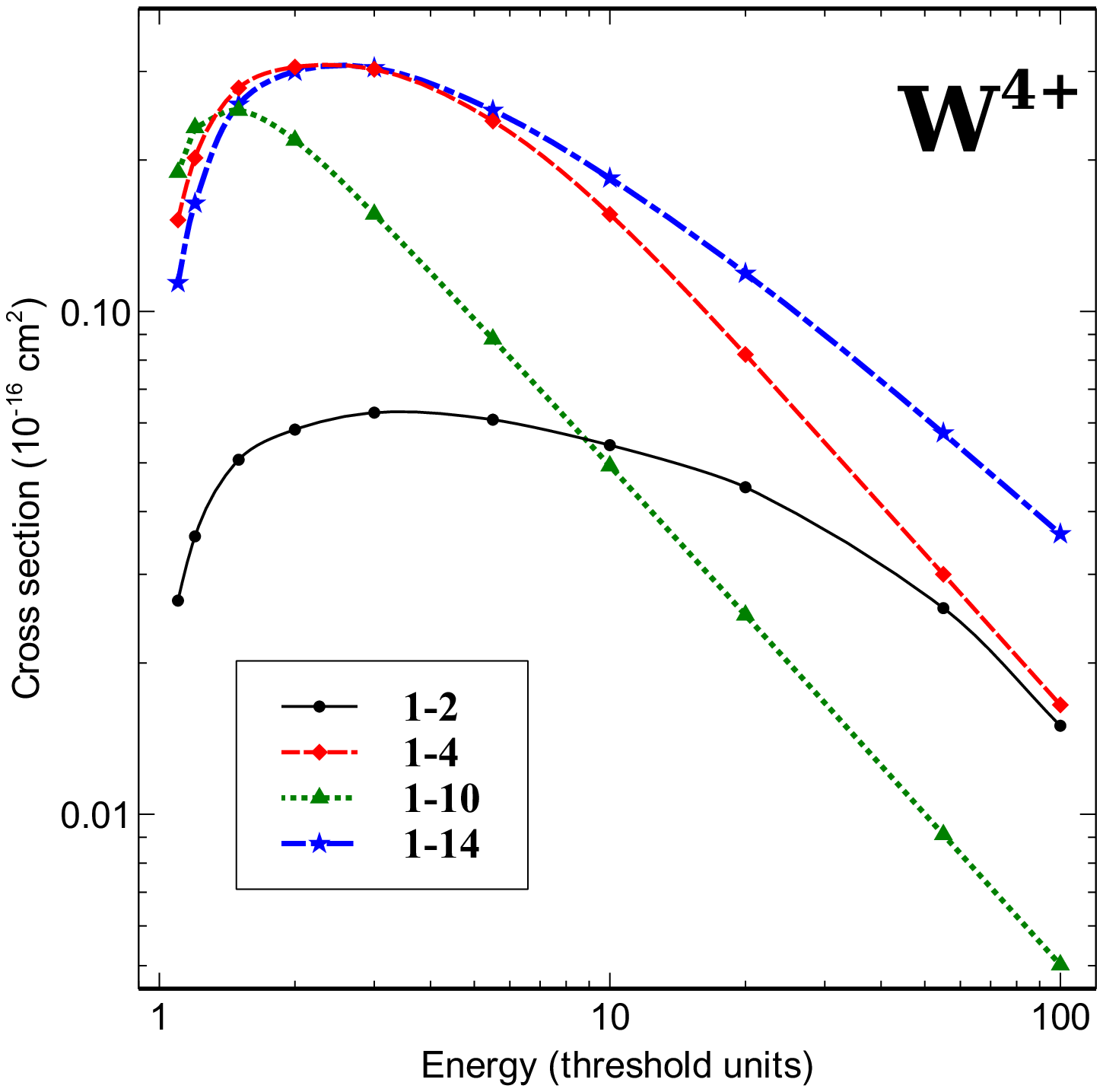}
\caption{\label{w4} 
Electron-impact excitation cross sections (in $10^{-16}$ cm$^2$) 
from the ground state 5p$^6$5d$^2$\,\,$^3$F$_{2}$ 
of the W$^{4+}$ ion to the excited states
5p$^6$5d$^2$\,\,$^3$F$_{3}$ (solid curve),
5p$^6$5d$^2$\,\,$^3$P$_{0}$ (dashed curve),
5p$^6$5d6s\,\,$^3$D$_{1}$ dotted curve) and 
5p$^6$5d6p\,\,$^3$F$_{2}$ (dash-dotted curve). 
Incident electron energies are given in transition threshold units. 
}
\end{figure}

\begin{figure}
\includegraphics[scale=0.5]{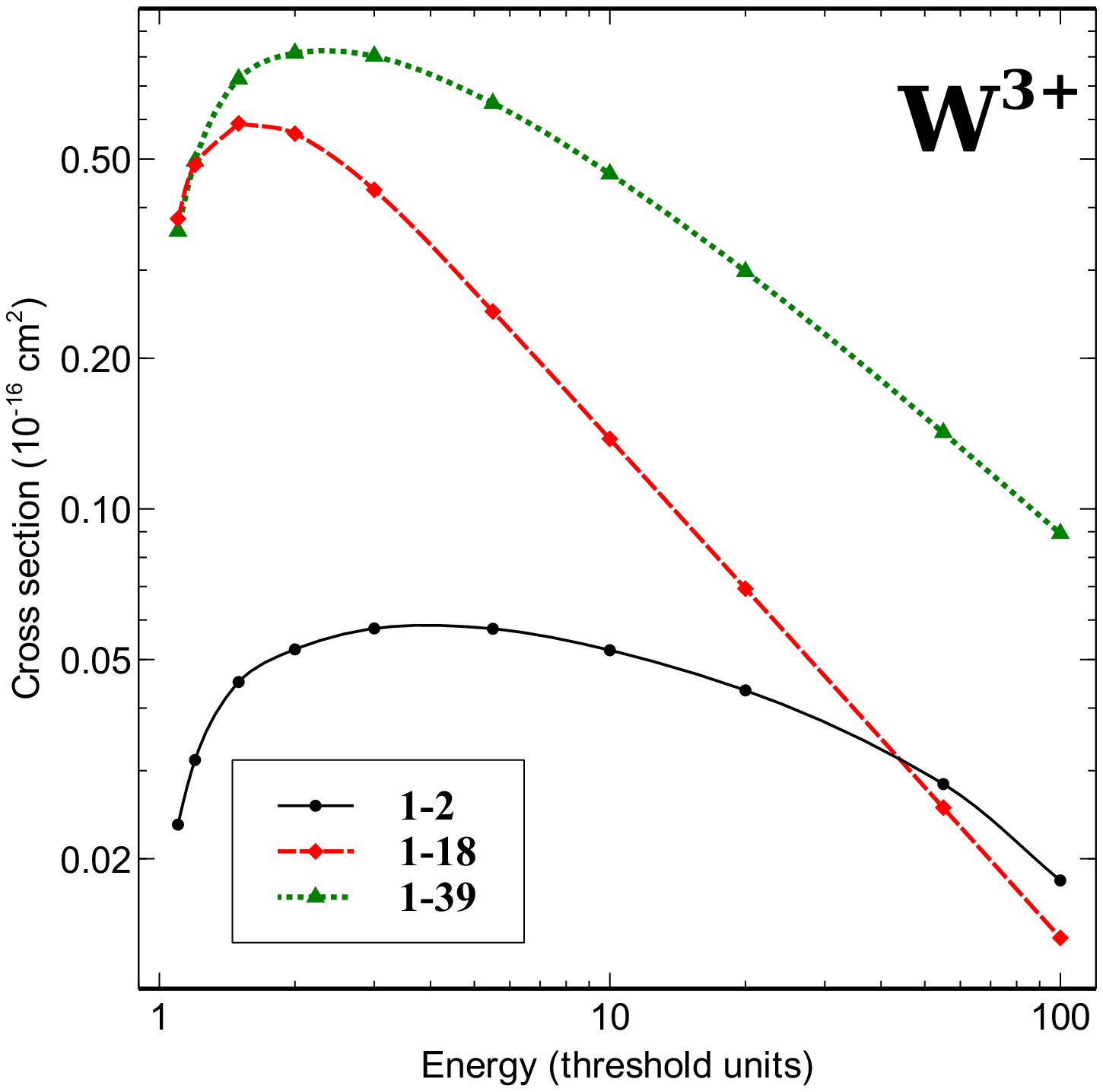}
\caption{\label{w3} 
Electron-impact excitation cross sections (in $10^{-16}$ cm$^2$)
from the ground state 5p$^6$5d$^3$\,\,$^4$F$_{3/2}$
of the W$^{2+}$ ion to the excited states
5p$^6$5d$^3$\,\,$^4$F$_{5/2}$ (solid curve),
5p$^6$5d$^2$6s\,\,$^4$F$_{3/2}$ (dashed curve) and 
5p$^6$5d$^2$6p\,\,$^4$F$_{3/2}$ (dotted curve).
Incident electron energies are given in transition threshold units. 
}
\end{figure}


\begin{figure}
\includegraphics[scale=0.5]{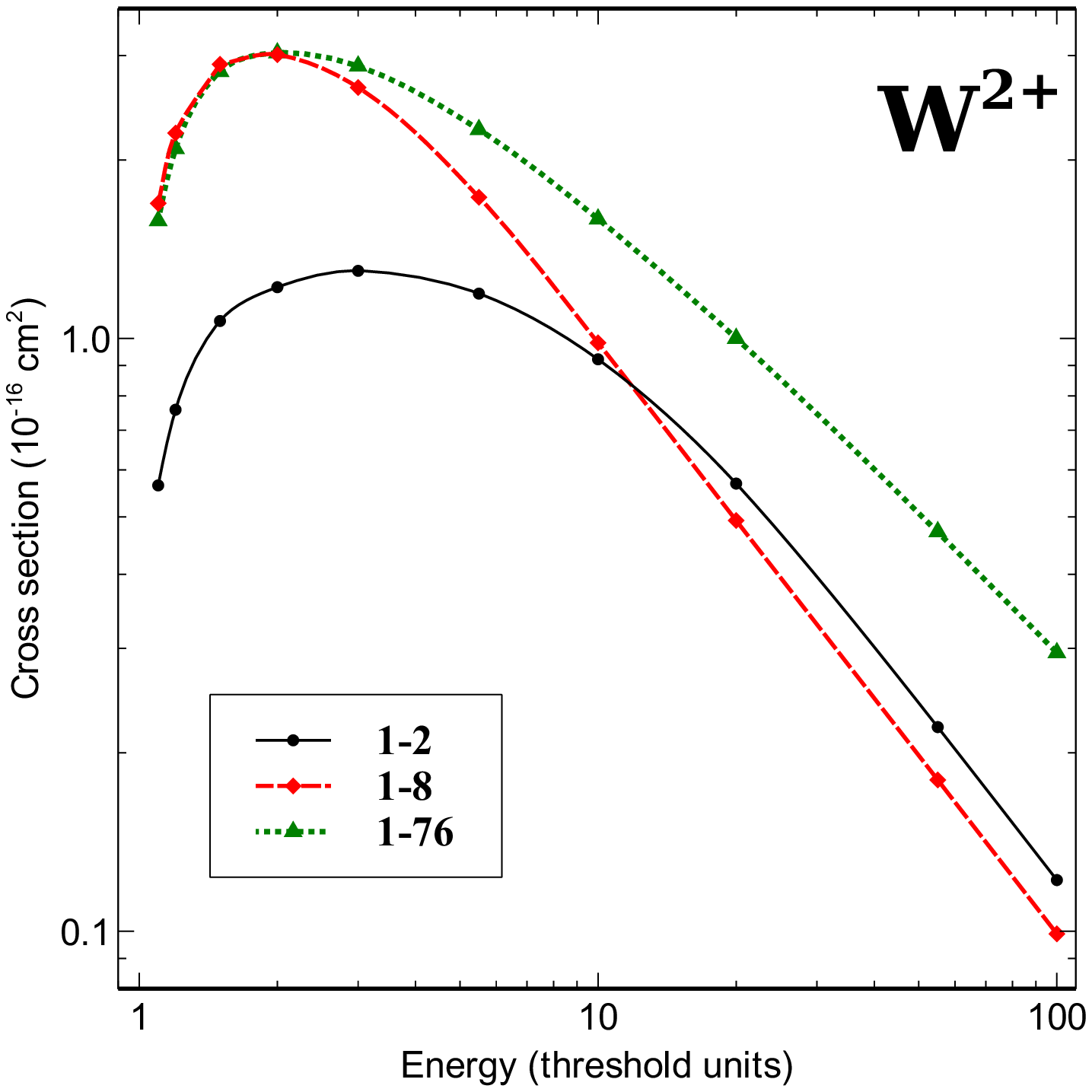}
\caption{\label{w2} 
Electron-impact excitation cross sections (in $10^{-16}$ cm$^2$) 
from the ground state 5p$^6$5d$^4$\,\,$^5$D$_{0}$ 
of the W$^{2+}$ ion to the excited states
5p$^6$5d$^4$\,\,$^5$D$_{1}$ (solid curve),
5p$^6$5d$^3$6s\,\,$^5$F$_{2}$ (dashed curve) and 
5p$^6$5d$^3$6p\,\,$^3$D$_{1}$  (dotted curve).
Incident electron energies are given in transition threshold units. 
}
\end{figure}

To demonstrate the abilities of our developed computer codes, we present several
plots of the electron-impact excitation cross sections for the tungsten ions
ranging from W$^{5+}$ to W$^{2+}$. Our performed calculation involves a large
number of excitations involving levels of these ions. All possible excitation
transitions involving several hundreds of levels are considered. Plots given in 
the present paper are given for demonstration purpose.

Figures \ref{w5_1} and \ref{w5_2} demonstrate cross sections for the W$^{5+}$ 
ion excitation from the ground and from the first excited level. For other three
ions, namely W$^{4+}$, W$^{3+}$ and W$^{2+}$, we present only the excitation 
parameters from the ground level in Figs. \ref{w4}, \ref{w3}, \ref{w2}, 
although much more data are produced. For example, we consider the lowest 293 
levels in W$^{2+}$ and transitions among them.

\section{Summary and conclusions}
\label{5}

The performed calculations of the electron-impact excitation data and their 
analysis have demonstrated that the computational methods implemented in our
newly-developed computer codes are suitable to produce the electron-impact 
excitation cross sections and related parameters, such as the electron-impact 
collision strengths or collision rates. One can expect that such the data for 
other ions will be reliable enough.

Based on good agreement of the radiative transition probabilities with 
existing data, we expect that our determined electron-impact excitation cross 
sections for the 2s--3p transition in Li atom are accurate enough. 
The comparison of cross sections for W$^{45+}$ leads to conclusion that
the calculated parameters even for highly-charged ions can be applied in
high-temperature fusion plasma spectra modelling. Significant deviations from
earlier published results can be explained by the fact that radiative transition
partameters are influenced by inclusion of correlation corrections differently 
compared to excitation cross sections. Therefore a simple normalization 
described by Eq.(\ref{sigmaBEF}) in \cite{8} is not well-grounded.

We are planning to incorporate results of such calculations (together with other
spectroscopic parameters) into the newly-developed database ADAMANT when the
electron-impact excitation processes in various many-electron atoms and ions
with open s-, p-, d- and f- shells are considered.

\ack
D.S. acknowledges support by project "Promotion of Student Scientific 
Activities" (VP1-3.1-{\v S}MM-01-V-02-003) from the Research Council of 
Lithuania, funded by the Republic of Lithuania and European Social Fund under 
the 2007 -- 2013 Human Resources Development Operational Programme's 
priority 3. P.B. and R.K. acknowledge support by NSF grant AST/1109061.

\end{document}